\documentclass[article]{aa}
\usepackage{graphicx,epsfig,epsf,rotating,fancyhdr,amsmath,natbib}
\usepackage{amsmath}
\usepackage{txfonts}

\begin{document}

\title{Modelling the radio pulses of an ultracool dwarf}
\author{S. Yu\inst{1}, G. Hallinan\inst{2,3}, J.G. Doyle\inst{1}, A.L. 
MacKinnon\inst{4}, A. Antonova\inst{5}, A. Kuznetsov\inst{1,6},
A. Golden\inst{2}, Z.H. Zhang \inst{7}}
\institute {Armagh Observatory, College Hill, Armagh BT61 9DG, N. Ireland\\
              \email{syu@arm.ac.uk}
\and Centre for Astronomy, National University of Ireland, Galway, Ireland
\and Department of Astronomy, University of California, Berkeley, CA 94720-3411,
USA
\and Department of Adult and Continuing Education, University of Glasgow, Scotland
\and Department of  Astronomy, Faculty of Physics, St
Kliment Ohridski, University of Sofia, Bulgaria
\and Institute of Solar-Terrestrial Physics, Irkutsk 664033, Russia
\and Centre for Astrophysics Research, Science and Technology Research Institute, University of Hertfordshrie, Hatfield AL10 9AB
}

\date{Received ; accepted }

\abstract
{Recently, unanticipated magnetic activity in ultracool dwarfs (UCDs, spectral
classes later than M7) has emerged from a number of radio observations. 
The highly (up to 100\%) circularly polarized nature and high brightness 
temperature of the emission have been interpreted as requiring an effective amplification 
mechanism of the high-frequency electromagnetic waves $-$ the electron cyclotron
maser instability (ECMI).}
{We aim to understand the magnetic topology and the properties of the radio 
emitting region and associated plasmas in these ultracool dwarfs, interpreting 
the origin of radio pulses and their radiation 
mechanism.}
{An active region model was built, based on the rotation of the UCD and the ECMI mechanism. }
{The high degree of variability in the brightness and the diverse profile of pulses can be
interpreted in terms of a large-scale hot active region with extended magnetic 
structure existing in the magnetosphere of TVLM 513-46546. 
We suggest the time profile of the radio light curve is 
in the form of power law in the model. Combining the 
analysis of the data and our simulation, we can 
determine the loss-cone electrons 
have a density in the range of $1.25\times10^{5}-5\times10^{5}$ cm$^{-3}$ and 
temperature between $10^{7}$ and $5\times10^{7}$ K. The active region 
has a size $<1~R_{\rm Jup}$, while the pulses produced by 
the ECMI mechanism are from a much more compact region (e.g. $\sim$0.007 $R_{\rm Jup}$). 
A surface magnetic field strength of $\approx$7000 G is predicted.}
{The active region model is applied to the radio emission from TVLM 513-46546, in 
which the ECMI mechanism is responsible for the radio bursts from the 
magnetic tubes and the rotation of the dwarf can modulate the integral of flux 
with respect to time. The radio emitting region consists of complicated substructures. 
With this model, we can determine the nature (e.g. size, temperature, density) 
of the radio emitting region and plasma. The magnetic topology can also be 
constrained. We compare our predicted X-ray flux with Chandra X-ray observation of TVLM 513-46546. 
Although the X-ray detection is only marginally significant, our predicted flux is significantly 
lower than the observed flux. Further multi-wavelength observations will help us better understand 
the magnetic field structure and plasma behavior on the ultracool dwarf.}

\authorrunning{Yu et al.}
\titlerunning{Modelling the radio pulses of an ultracool dwarf}

\maketitle

\section{Introduction}
\label{intro}

A large number of recent radio observations indicate that intense magnetic
activity exists in ultracool dwarfs (UCDs), i.e. objects with spectral type 
later than M7. \citet{Berger01} reported the first detection of quiescent 
and flaring radio emission from the M9 brown dwarf LP944-20, in which a bright
X-ray flare was also detected \citep{Rutledge00}, with anomalous quiescent 
radio luminosity at least four orders of magnitude larger than predicted from an 
empirical relation between the X-ray and radio luminosities of active stars 
with spectral types from F to M \citep{Gudel93}. The detection of electron 
cyclotron maser (ECM) emission provided 
the first confirmation of kilogauss fields for a late M dwarf \citep{Hallinan06}, 
and subsequently led to the discvery that even cooler L type dwarfs 
can also possess magnetic fields in the kilogauss range \citep{Hallinan08}. 
Radio observations exclusively enable the measurement of 
magnetic fields on cool, brown dwarfs, and possibly also for the very 
faint exoplanets \citep{Zarka07}.

Chromospheric H$\alpha$ emission and coronal X-ray emission show  
a sharp decline in $L_{\rm H\alpha}/L_{\rm bol}$ and $L_{\rm X}/L_{\rm bol}$ 
beyond spectral type M7 \citep{Neuhauser99,Gizis00,West04,Stelzer06a,Schmidt07} 
which would be consistent with lower fractional ionization in  
atmospheres of later spectral type \citep{Mohanty02}. 
Recently, however, a lot of evidence, such as quiescent and flaring H$\alpha$ 
emission from some L and T dwarfs \citep{Reid99,Burgasser00,liebert03,Reiners07,
Rockenfeller06,Stelzer06b,Schmidt07}, FeH lines from cool M dwarfs 
\citep{Afram09}, strong X-ray emission from one such source  
\citep{Audard07}, all suggest that perhaps an efficient magnetic dynamo could 
be operational \citep{Parker55} in a fraction of the UCD population 
\citep{Reiners10} and that magnetic 
reconnection events could occur on these kinds of cool objects. Such magnetic 
activity and thus the strong radio emission could be associated with 
the differential rotation between the atmosphere and the core of the UCD.

\citet{Berger02} reported Very Large Array (VLA) observations of 12
late M and L dwarfs in the solar neighborhood. Flare-like outbursts as well 
as persistent quiescent emission were detected from three of the 12
sources, TVLM 513-46546 (TVLM 513 hereafter), 2MASS J00361617$+$1821104 (2MASS J0036$+$18) and BRI 0021-0214. 
Among these three radio-active sources, TVLM 513 and BRI 0021, plus the first detected
radio-active UCD, LP944-20, all have rapid rotational velocities,
$v\sin i>$30 km s$^{-1}$. \citet{Hallinan06} presented
observations of the rapidly rotating M9 dwarf TVLM 513
obtained simultaneously at 4.88 and 8.44 GHz using the VLA. The periodic radio
emission at both frequencies indicated a period of $\sim$2 hr in
excellent agreement with the estimated period of rotation of the
dwarf based on its $v\sin i$ of $\sim$60 km s$^{-1}$. One more
radio-active source from \citet{Hallinan08}, LSR J1835$+$3259,
also has a high rotational velocity, $v\sin i\approx$50 km s$^{-1}$.

Up to now, about 10 radio active
ultracool dwarfs, with spectral type from M8 to L3.5, including a
binary system (2MASS J07464256+2000321, hereafter 2MASS 0746$+$20), have been found from various
surveys \citep{Berger02,Burgasser05,Berger06,Hallinan06,Hallinan07,Hallinan08,
Antonova08}. Three of these have been shown to have
periodic radio emission, with periods of 1.96 hr (TVLM 513, 
\citet{Hallinan07}), 3.07 hr 
(2MASS J0036$+$18, \citet{Hallinan08}), and 2.83 hr (LSR 1835$+$32, 
\citet{Hallinan08}). 

The L dwarf binary, 2MASS 0746$+$20, reported by \citet{Antonova08} 
from a mini-survey of UCDs at 4.9 GHz has a high mean flux level of 
286$\pm$24 $\mu$Jy. \citet{Berger09} presented an 8.5 hr simultaneous 
radio, X-ray, UV, and optical observation of this binary. 
The strong radio emission consists mainly of short-duration periodic pulses 
at 4.86 GHz with $P$ $=$ 124.32 $\pm$ 0.11 min. The radio pulses are 1/4 phase 
different from the H${\alpha}$ emission.

The narrow bunching of multiple pulses of both left- and 
right- 100\% polarized radio emission detected from TVLM 513, which originate 
in regions of opposite magnetic polarity, reveal the likely 
presence of a dipolar component to the large-scale magnetic field \citep{Hallinan08}. 
Zeeman Doppler Imaging (ZDI) observations have also shown that such a large-scale dipolar magnetic 
structure could exist on an M4 dwarf star, V374 Peg, which is also a fully 
convective rapid rotator similar to TVLM 513 \citep{Donati06,Morin08}. 
The topology of magnetic fields on UCDs needs to be constrained by more 
observations.

The radio emission composed of quiescent and pulsing components from UCDs 
can be associated with not only the geometry of the emitting region and rotation 
of UCDs, but also the behavior of plasma in the magnetic field, i.e. the radiation 
mechanism. The very high brightness temperature and high (up to 100\%) circular 
polarization of the pulses \citep{Hallinan07} point towards an efficient,  
coherent radiation mechanism, the electron cyclotron maser instability (ECMI, 
\citet{Melrose82}). This wave magnification process of the free-space radiation 
modes could be induced by some kind of anisotropic velocity distributions of electrons, 
such as a loss-cone distribution \citep{Lau83}, ring shell distribution, or 
horseshoe distribution \citep{Pritchett84}. 

ECMI was successfully applied to the auroral kilometric radiation (AKR) on Earth 
\citep{Wu79,Ergun00}, 
decametric radiation (DAM) on Jupiter, Saturnian kilometric radiation (SKR) 
\citep{Zarka98,Zarka04} and solar millisecond microwave spikes 
\citep{Aschwanden90b}. Various authors have suggested its presence in exoplanets 
\citep{Zarka07,Griebmeier07,Jardine08}. 
The magnitude of any contribution from incoherent gyrosynchrotron or synchrotron 
radiation to the quiescent components in UCDs is uncertain.

The generation of ECMI emission is dependent on the environment of the 
emitting region, such as the magnetic field (strength, structure), the 
electron density distribution (number, energy), the line-of-sight source scale, 
and the angle between the line of sight and the magnetic field \citep{Melrose82}, 
and also the details (velocity space gradients) of the loss cone distribution \citep{Aschwanden90a}. 
This results in the possibility of transient radio emission. 
The findings of \citet{Antonova07} indicate that UCDs may 
also have sporadic long-term variability in
their levels of quiescent radio activity. This phenomenon could be
related to the change of the environment of the radio-emitting region, leading to 
self-quenching of the electron-cyclotron maser. Radio observations can help us 
to determine the magnetic configuration and the nature of the plasma in the emitting region.

For the purpose of understanding the magnetic topology, we 
built an active region model based on the rotation of TVLM 513 and the ECMI mechanism 
to simulate the observed light curve. We summarize the previous radio observation 
on TVLM 513 in \S\ref{sec_tvlm}. In \S\ref{sec_model}, we present the model, 
and results are given in \S\ref{sec_results}. We discuss these results  
in \S\ref{sec_discussion} and draw conclusions in \S\ref{sec_conclusions}.

\section{Previous observations on TVLM 513}
\label{sec_tvlm}

TVLM 513 is a young radio active M8.5V dwarf with a bolometric magnitude of 
log$\left(L_{\rm bol}/L_{\odot}\right)\approx-3.65$, effective temperature 
$T_{\rm eff}\approx2200$K \citep{Tinney93,Tinney95,Leggett01}, and situated at a distance of 
$d=10.6$ pc \citep{Dahn02}. From the theory of the formation 
and evolution of UCDs and the absence of lithium on TVLM 513, it is 
reasonable to infer values for the mass and radius of this star of $\sim$0.07 
$M_{\odot}$ and $\sim$0.1$R_{\odot}$ respectively \citep{Reid02,Chabrier00}.

With the VLA a highly right-circularly polarized ($\sim$65\%) radio event from TVLM 513 was 
detected with a flux density of $\sim$1100 $\mu$Jy, as well as persistent variable 
emission at 8.46 GHz (\citet{Berger02}). \citet{Osten06} 
conducted a multifrequency VLA observation of TVLM 513 at 8.4, 4.8 and 1.4 GHz, 
using a strategy that involved time-sharing a single 10 hr observation between the 
various frequency bands. TVLM 513 was detected at each frequency band with only marginal 
confirmation of variability and no detection of flares or strong circular polarization.

Again, using the VLA, \citet{Hallinan06} found persistent and periodic 
radio emission from TVLM 513 at 8.44 GHz and 4.88 GHz simultaneously, 
with a period of $\sim$2 hr.
Subsequently, extremely regular periodic bursts (p $=$ 1.96 hr, up to $\sim$4 mJy) 
of high brightness and highly circularly polarized radio emission were 
reported by \citet{Hallinan07}. 
Multiple bursts of both left and right 100\% circularly polarized emission were 
detected. Interestingly, the radio emission can switch states from nearly 100\% left 
polarization to 100\% right polarization in each phase. 

Another radio burst with a flux density up to $\sim$4 mJy was presented  
by \citet{Berger08} from a period of simultaneous radio, X-ray, ultraviolet, and optical spectroscopic 
observations. Steady quiescent radio emission superposed with 
multiple, short-duration, highly polarized bursts was observed, but these authors reported 
a non-periodicity in the pulses/flaring activity. In a re-analysis of this data, plus data 
taken $\approx$40 days later (June 2007), \citet{Doyle10} reported the 1.96 hr periodicity in both
datasets, deriving a more accurate period. 
\begin{figure}[ht!]
\centering
\includegraphics[width=9cm,clip]{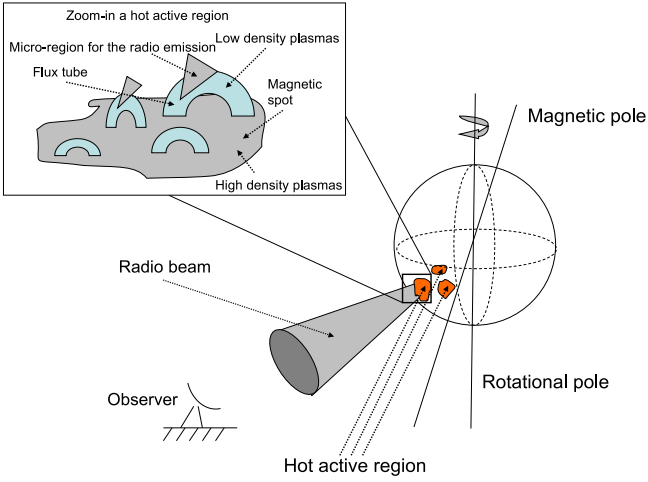}
\caption{A sketch of the active region model on ultracool dwarfs.}
\label{fig_ucdspot}
\end{figure}

\section{Model}
\label{sec_model}

In this section, we present a model (Figure \ref{fig_ucdspot}) to simulate the observational time profile 
of the radio flux density from TVLM 513 by assuming ECM emission is the dominant 
radiation mechanism. 

When plasma electrons are energized in magnetic flux 
tubes with converging legs and foot-points in a high density atmosphere (perhaps as 
a consequence of magnetic reconnection), some of these fast electrons collide with the high density 
atmosphere and thermalise at the foot-points. The remaining fast electrons are reflected in 
the converging field by a magnetic mirror effect. This process results in the formation of 
the anisotropic distribution of the plasma in velocity space, i.e. a loss-cone distribution. 

The plasma including a loss-cone distribution is unstable; instability arises 
very quickly from such a distribution. A large amount of free energy can be released
via the instability and converted to electromagnetic waves - see \citet{Dulk85} for a review of 
the process. An external magnetic-field-aligned electric field 
\citep{Cattell98,Zarka98,Ergun00} induced by a time-varying external current source \citep{Omura03} 
would further modify the plasma velocity distribution to shell or 
horseshoe form, leading to an enhanced ECMI emission. In this paper, we assume for simplicity that 
the ECMI emission from UCDs is driven by the loss-cone distribution. The existence 
of the electric field and its effect will be addressed in future work.

In a loss-cone region where the ECMI operates, the maser radiation is concentrated on the 
surface of a hollow cone as discussed by \citet{Melrose82} (see Fig. \ref{fig_ucdspot} top panel). 
The half-angle $\alpha_{0}$ of the hollow cone depends on the ratio of the velocity of the plasma electrons 
to the speed of light, i.e. cos$\alpha_{0}=v_{p}/c$. For example, if $v_{p}/c=0.5$, we have 
$\alpha_{0}=60^{\circ}$. The surface of the cone should be very thin with $\Delta\alpha\approx v_{p}/c$. 
The maser emission in the loss-cone region can be observed if the line of sight is located within a thin 
conical sheet with thickness $\Delta\alpha$. For a low magnetic loop with a small angle between the 
magnetic field and the surface of the UCD, the maser emission can be seen when the emission is near the top 
of the loop. The maser emission from near the foot-point (where the magnetic field is almost 
perpendicular to the surface of UCD) can be seen when the loop is near the limb. However, for a large 
scale, the maser emission could have an angular distribution. In our calculation, we assume we can observe 
the maser emission in each flux tube in the active region. 

By analogy to the solar coronal radio emission powered by two populations of plasma 
from the Sun with different velocity, we propose there be similar active regions  
on UCDs. We note that, however, an alternative mechanism, 
where the hot plasma beam could result from the interaction of a close-in companion of the UCD, 
i.e. magnetized or non-magnetized satellites resulting in auroral emission, 
similar to Io-Jupiter system \citep{Queinnec98,Saur04,Zarka05}, 
is possible. ECM emission has 
been detected in compact objects, such as white dwarfs \citep{Willes04,Willes05} 
or neutron stars \citep{Wolszczan92}. We are not able to rule out this model and 
the competition between the two models should be investigated in further work.

\begin{figure*}[ht!]
\hspace*{-1cm}
\includegraphics[width=20cm,clip]{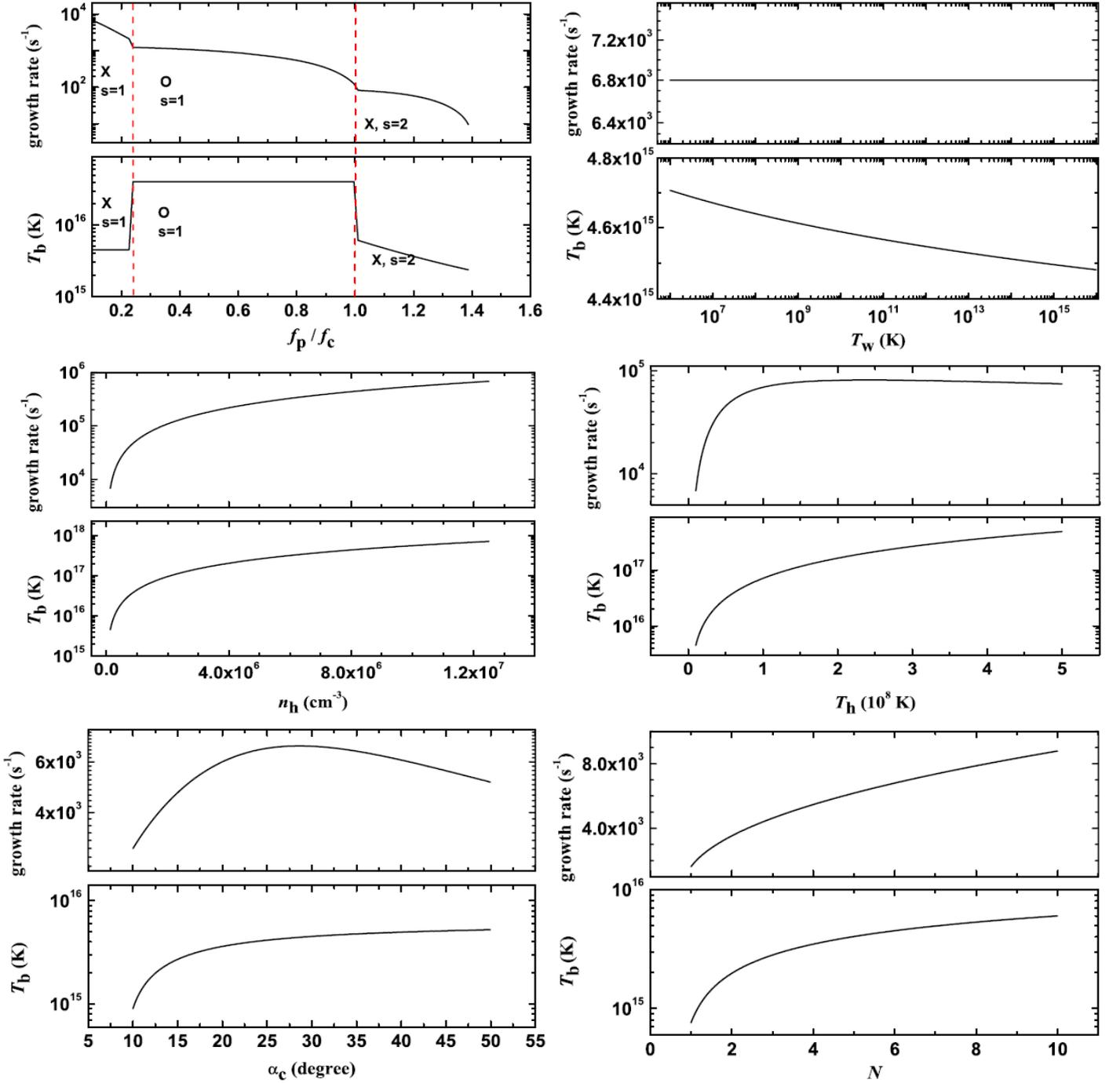}
\caption{The change of growth rate and brightness temperature with different plasma parameters 
and loss cone parameters. $f_{\rm p}$/$f_{\rm c}$ = ratio of plasma frequency and gyrocyclotron frequency, 
$T_{\rm w}$ = background wave energy, $n_{\rm h}$ = density of hot plasma, $T_{\rm h}$ = temperature of hot plasma, 
$\alpha_{\rm c}$ = loss cone angle, $N$ = loss cone pitch angle distribution slope, $T_{\rm b}$ = brightness temperature. 
The range of each parameter is in the definition of \citet{Aschwanden90a}. We plot each panel assuming 'standard' values of 
the other parameters (see Table \ref{tab_parameters}).}
\label{fig_grbt}
\end{figure*}

\subsection{Flux calculation}

The fine structure and time interval of the observations described in \S \ref{sec_tvlm} 
indicate that the radio-emitting region on TVLM 513 
would consist of complex substructures. 
The detected flux density $\tilde{S}$ may be the 
sum of several small sources 
\begin{equation}
\tilde{S}=\sum_{\rm i=0}^{\tilde{n}(t)}S_{\rm i}.
\end{equation}
where $S_{\rm i}$ is the flux density for a small source which can be determined by the relation \citep{Dulk85}
\begin{equation}
S_{\rm i}=\int k_{\rm B} \left(\frac{f}{c}\right)^{2} T_{\rm b} d\Omega
\end{equation}
where $T_{\rm b}$ is the brightness temperature of a source, $k_{\rm B}$ is the Boltzmann constant, 
$f$ is the observed frequency, $c$ is the speed of light, d$\Omega$ is the differential solid angle. 
If we assume that the radiation is isotropic, the differential solid angle should depend on the 
radius of the flux tube $r_{\rm tube}$ of the small source and its distance from the observer. 
The flux density $S_{i}$ can be expressed as 
\begin{equation}
\label{eq_fluxtube}
S_{\rm i}= k_{\rm B} \left(\frac{f}{c}\right)^{2} T_{\rm b} \frac{\pi r_{\rm tube}^{2}}{4\pi d^{2}}.
\end{equation}

For simplicity, we assume that the active region has a symmetric shape, and the small sources are randomly
distributed within the region. The number of small sources $\tilde{n}(t)$ may be determined by rotation 
of the UCD. We set the time $t=0$ to be the moment when the first active region emerges in the field of view of the 
observer; this is also the onset time of a radio pulse. With the rotation of the UCD, the area of 
the active region seen by the observer increases until it reaches a maximum, and then it decreases and 
disappears from the field of view of the observer. The maximum number of small sources is
\begin{equation}
\tilde{n}_{max}\approx \left(\frac{\Delta t \cdot 2\pi R' \cos\theta/T_{\rm UCD}}{r_{\rm tube}}\right)^{\gamma}.
\label{eq_maxnum}
\end{equation} 
where $T_{\rm UCD}$ is the rotation period, $R'$ is the height of the active region, $\theta$ is the latitude 
of the active region and $\Delta t$ is the time interval from the beginning of a pulse to the 
maximum flux. The time duration of a pulse would be 2$\Delta t$ for a symmetric shape. We assume that the radio 
emission is from a thin shell of the active region near the surface of the UCD, so that we can take 
$\gamma\approx2$ and $R'\approx R_{\rm UCD}$ where $R_{\rm UCD}$ is the radius of the UCD.

\begin{table}
\begin{minipage}[t]{\columnwidth}
\caption{Parameters to determine the flux density. Plasma parameters: 
$n_{\rm h}$ = density of hot plasma, $T_{\rm h}$ = temperature of hot plasma, 
$u=f_{\rm p}/f_{\rm c}$ = ratio of plasma frequency and gyrocyclotron frequency,
$n_{\rm c}$ = density of cold (background) plasma, $T_{\rm c}$ = temperature of 
cold (background) plasma, $T_{\rm w}$ = background wave energy; Loss cone parameters: 
$\alpha_{\rm c}$ = loss cone angle, $N$ = loss cone pitch angle distribution slope, $B$ = magnetic field 
strength; UCD parameters: $R_{\rm UCD}$ = radius of UCD, $T_{\rm UCD}$ = rotation 
period of UCD, $r_{\rm tube}$ = radius of flux tube, $\theta$ = latitude of 
radio-emitting region, $A$ = size of emitting region. Note that the number density 
of cold plasma is associated with $f_{\rm p}$/$f_{\rm c}$.}
\label{tab_parameters}
\begin{center}
\begin{tabular}{lccccccccc}
\hline
Plasma
 & $n_{\rm h}$                   & $T_{\rm h}$                & $u~(f_{\rm p}$/$f_{\rm c})$ & $T_{\rm c}$ & $T_{\rm w}$  \\
 & (cm$^{-3}$)                   & (K)                        &                         &  (K)        &  (K)  \\
 & 1.25$\times10^{5}$             & $10^{7}$                   &0.1                      & $10^{6}$    & $10^{14}$\\
 \hline  
Loss-cone
 & $\alpha_{c}$ &  $N$  & $B$  \\
 & (degree)     &       & (G) \\
 & 30           &   6   & 1750 \\
 \hline 
UCD
 & $R_{\rm UCD}$                   & $T_{\rm UCD}$                & $r_{\rm tube}$ & $\theta$   &  $A$    \\
 & (km)                            & (hr)                        &  (km)          &   (degree) &     (km$^{2}$)   \\
 & 7.1$\times10^{4}$              & 1.96                          & 55             &     30     &   820$^{2}$      \\
\hline
\end{tabular}
\end{center}
\end{minipage}
\end{table}

\subsection{Brightness temperature and relative parameters}

The brightness temperature of small sources depends strongly on the growth rate 
of the ECM emission and incoherent radiation of the background plasma. If we assume 
the maser emission is operated by the loss-cone distribution of a population of hot 
plasma expressed by a Maxwellian multiplied by the function 
$\sin^{N}(\alpha/\alpha_{\rm c}\cdot \pi/2)$ and an isotropic Maxwellian 
distribution of a cold background plasma, the brightness temperature can be 
determined by the parameters of these two types of plasma. 
We adopt the quasi-linear theory developed by \citet{Aschwanden90a} to determine 
the growth rate and the efficiency of energy conversion. Here, we summarize the 
basic assumptions and the effects of several free parameters. 

The quasi-linear code of \citet{Aschwanden90b} describes the evolution of the ECM 
instability and the wave-particle interactions by solving the kinetic wave-particle 
equations in a locally homogeneous plasma. The wave equation includes induced 
gyroresonance emission/absorption (for the X-, O-, Z- magneto-ionic modes and whistlers), 
but neglects spontaneous emission, free-free absorption, collisional deflection, and 
spatial wave propagation. The coupled diffusion equation contains the quasi-linear 
diffusion coefficients due to maser growth/damping, but neglects slower processes 
like particle loss and source terms. This assumption corresponds to the strong diffusion 
case. The quasi-linear diffusion process of the ECM instability in the solar corona 
successfully accounted for the time profile and observational characteristics of decimetric 
millisecond spikes \citep{Aschwanden90b}. 

In order to constrain the growth rate $\Gamma$, energy conversion factor 
$\varepsilon_{\rm c}$ and saturation of the maser, we need to know the initial 
nature of the two populations of plasma and the shape of the loss-cone, i.e. 
for the hot plasma: particle density $n_{\rm h}$, particle temperature $T_{\rm h}$; 
for the cold plasma: particle density $n_{\rm c}$, particle temperature $T_{\rm c}$; 
for the loss-cone: loss-cone angle $\alpha_{\rm c}$, pitch angle distribution slope $N$. 
Rather than choowing values of the cold plasma density $n_{\rm c}$ we fix its value via adoption
of a value for $u=f_{\rm p}/f_{\rm c}$, the ratio 
of plasma frequency ($f_{\rm p}\simeq 9\times10^{-3}(n_{\rm c})^{1/2}$ MHz) 
and gyrofrequency ($f_{\rm c}\simeq2.86\times B$ MHz) where $B$ is the magnetic field strength in G.

In the quasi-linear diffusion process, the existence of background electromagnetic wave energy $T_{\rm w}$ 
is taken into account. As discussed by \cite{Aschwanden90a} in the case of solar millisecond spikes, 
the initial brightness temperature 
could be as low as the level of thermal bremsstrahlung in the range of $10^{6}\sim10^{8}$ K. However, 
because of gyroresonance or gyrosynchrotron radiation, an enhanced photon level could exist in a flaring 
loop which would affect the ECM process significantly, yielding a wave turbulence at the level of 
$\sim10^{15}$ K. We apply $T_{\rm w}$ in the range of $10^{\rm 6}-10^{16}$ K in our model.

The free energy of the plasma with a loss-cone distribution can be 
converted into an equivalent electromagnetic energy. The energy conversion factor $\varepsilon_{\rm c}$ can 
be defined as the ratio of the change of kinetic energy between the initial state and the final state of 
the plasma and the initial kinetic energy. In the present paper, the amount of converted energy is 
about 0.5\%, i.e. the same as in \citet{Aschwanden90a}.  

\citet{Aschwanden90a} investigated how the parameters influence the growth rate and brightness 
temperature generated by the ECM instability and gives the general formula for the key parameters as 
follows:

\begin{equation}
\begin{split}
\Gamma&=6.9\times10^{5}\left(\frac{n_{\rm h}}{1.25\times10^{6}\rm cm^{-3}}\right)
\left(-1.1+1.1\left(\frac{\alpha_{\rm c}}{30^{\circ}}\right)+\left(\frac{\alpha_{\rm c}}{30^{\circ}}\right)^{-1}\right)^{-1}\\
&\times\left(0.65+0.05\left(\frac{T_{\rm h}}{10^{8}\rm K}\right)+0.30\left(\frac{T_{\rm h}}{10^{8}\rm K}\right)^{-1.5}\right)^{-1}
\left(1.15\left(\frac{N}{6}\right)-0.15\right)^{0.45}\\
&\times \begin{cases}
\left(2.0-1.2\left(\frac{u}{0.1}\right)+0.2\left(\frac{u}{0.1}\right)^{2}\right), ~~~~~~~~~~~~~~~~~~~~~~~~~~0.1<u\leqslant0.24\\
0.19(1.0092-0.0092\left(\frac{u}{0.1}\right)^{2}), ~~~~~~~~~~~~~~~~~~~~~~~~~0.24<u\leqslant1.0\\
0.021(-1.416+0.440\left(\frac{u}{0.1}\right)-0.024\left(\frac{u}{0.1}\right)^{2}),~~~ 1.0<u<1.4
\end{cases}
\end{split}
\end{equation}

\begin{equation}
\begin{split}
T_{\rm b}&=9.0\times10^{17}\left(\frac{n_{\rm h}}{1.25\times10^{6}\rm cm^{-3}}\right)^{1.10}
\left(1.4-0.4\left(\frac{\alpha_{\rm c}}{30^{\circ}}\right)^{-1}\right)\\
&\times \left(\frac{T_{\rm h}}{10^{8}\rm K}\right)^{1.2}\left(2.0\left(\frac{N}{6}\right)^{0.3}-1.0\right)\left(\frac{\log T_{\rm w}}{14}\right)^{-0.05}\\
&\times  \begin{cases}
1, ~~~~~~~~~~~~~~~~~~~~~~0.1<u\leqslant0.24\\
9, ~~~~~~~~~~~~~~~~~~~~~~0.24<u\leqslant1.0\\
1400\left(\frac{u}{0.1}\right)^{-3}, ~~~~1.0<u<1.4 
\end{cases}
\end{split}
\end{equation}
We plot in Figure \ref{fig_grbt} the effect of various parameters on the growth 
rate and brightness temperature.

\section{Results and comparison with observations}
\label{sec_results}

Many parameters, not only those describing the nature of the plasma but also those associated 
with the properties of the UCD, can affect the observable flux density significantly. In 
this section, we will discuss these parameters and compare our simulations with 
observations. In order to study the influence of the various physical quantities we adopt 
`standard' values for each parameter, listed in Table \ref{tab_parameters}. 
Unless otherwise noted, the simulations are always for magneto-ionic X-mode 
$\sigma=-1$, harmonic number $s=$1.

\subsection{The effect of rotation}
\label{sec_rotation}

In order to see the effect of rotation of the UCD, we first start a set of simulations by 
fixing the initial plasma and loss-cone parameters as listed in Table \ref{tab_parameters}. 
In addition, as suggested by \citet{Chabrier00}, we adopt the radius ($R_{\rm UCD}$
=0.1 $R_{\odot}$) as a constant in our simulations. We initially assume that the radio-emitting 
region is close to the equator (latitude $\theta=30^{\circ}$). 

\begin{figure}
\hspace*{-1cm}
\includegraphics[width=10cm,clip]{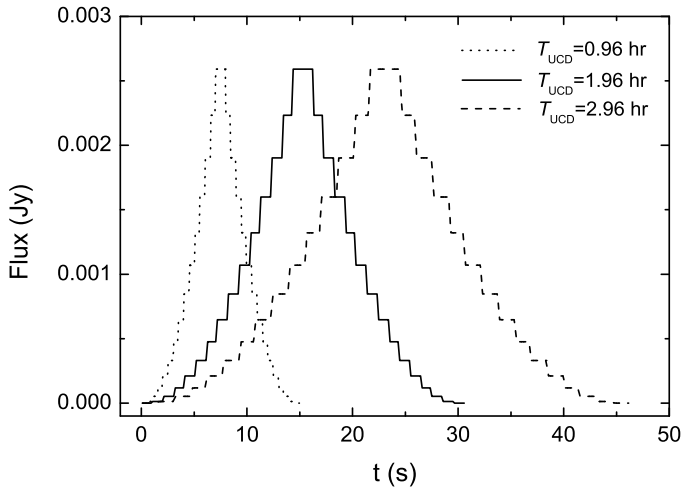}
\caption{The influence of rotation period ($T_{\rm UCD}$) on the radio light curve. 
Vertical scale denotes the flux density while the horizontal scale indicates the time. Dashed line, 
solid line and dotted line are for $T_{\rm UCD}$ 2.96, 1.96, 0.96 hr, respectively.}
\label{fig_tucd}
\end{figure}

Given the above parameters, the size ($A$) of the emitting region and the tube size 
of a small radio area, we show in Fig. \ref{fig_tucd} the influence of rotation. From the 
figure, we can see that 
the radio light curve is broadened with increasing rotation period ($T_{\rm UCD}$) 
while the intensity does not change at all. This is because the intensity of the radio emission 
is only related to the behavior of plasma and the total size of the emitting region.

\begin{figure*}
\centering
\includegraphics[width=16cm,clip]{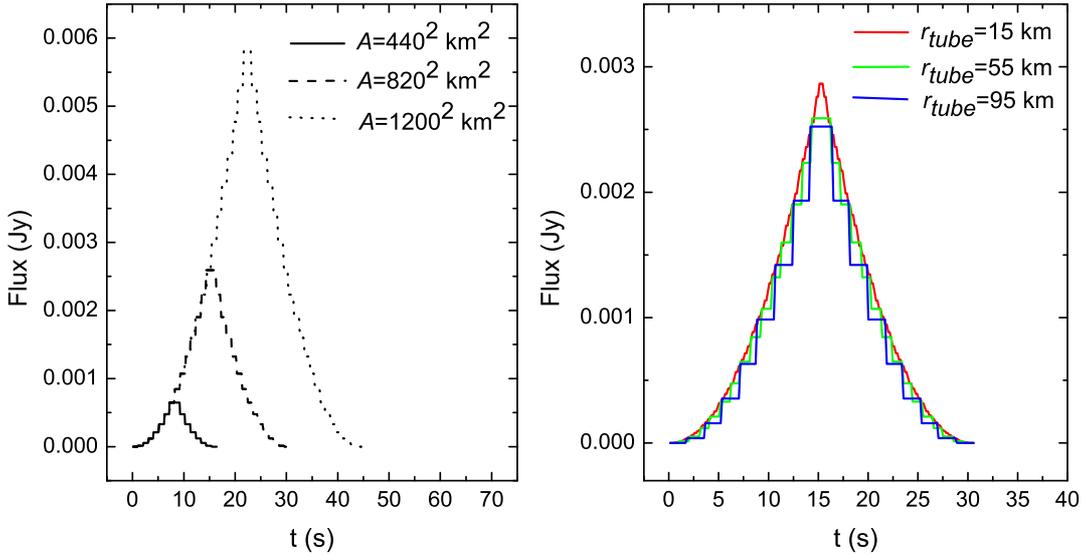}
\caption{Flux density against time with different size of the emission region ($A$, left 
panel) and radius ($r_{\rm tube}$, right panel) of one flux tube.}
\label{fig_rtubev}
\end{figure*}

Figure \ref{fig_rtubev} shows the effect of the total size ($A$) of the radio-emitting region 
(left panel) and the radius ($r_{\rm tube}$) of the flux tube (right panel). It is easy to 
understand that $r_{\rm tube}$ cannot change the radio flux as much as $T_{\rm UCD}$. It could, 
however, be a controlling factor for the observed oscillation of radio flux because each 
flux tube could have a different environment so that we may have to give a distribution 
for the initial plasma parameters (see \S \ref{sec_environment}).

$A$ is an important parameter for the nature of the radio-emitting region on the UCD, since we can easily 
see from Fig. \ref{fig_rtubev} that increasing its value can lead to a rise in both intensity 
and pulse duration. $A$ can be constrained by the observations once we know the rotation period 
of the UCD. We are able to evaluate $A$ from

\begin{equation}
A \sim (\Delta t \cdot 2\pi R_{\rm UCD} \cos\theta/T_{\rm UCD})^{\gamma}
\end{equation}
where the parameters are the same as in Eq. \ref{eq_maxnum}. For the standard parameters and $\gamma=2$, 
we find $\Delta t \approx 15$ s. This implies that the observed $\Delta t$ can be used to 
estimate $A$. 

\subsection{The effect of hot plasma in the loss-cone}

Because of the 
conservation of energy, the electromagnetic wave energy escaping from the radio-
emitting region has come from the kinetic energy of the hot plasma. So the 
brightness temperature of one flux tube should be proportional to the 
number density and temperature of hot particles. 

\begin{figure*}
\hspace*{-1cm}
\includegraphics[width=20cm,clip]{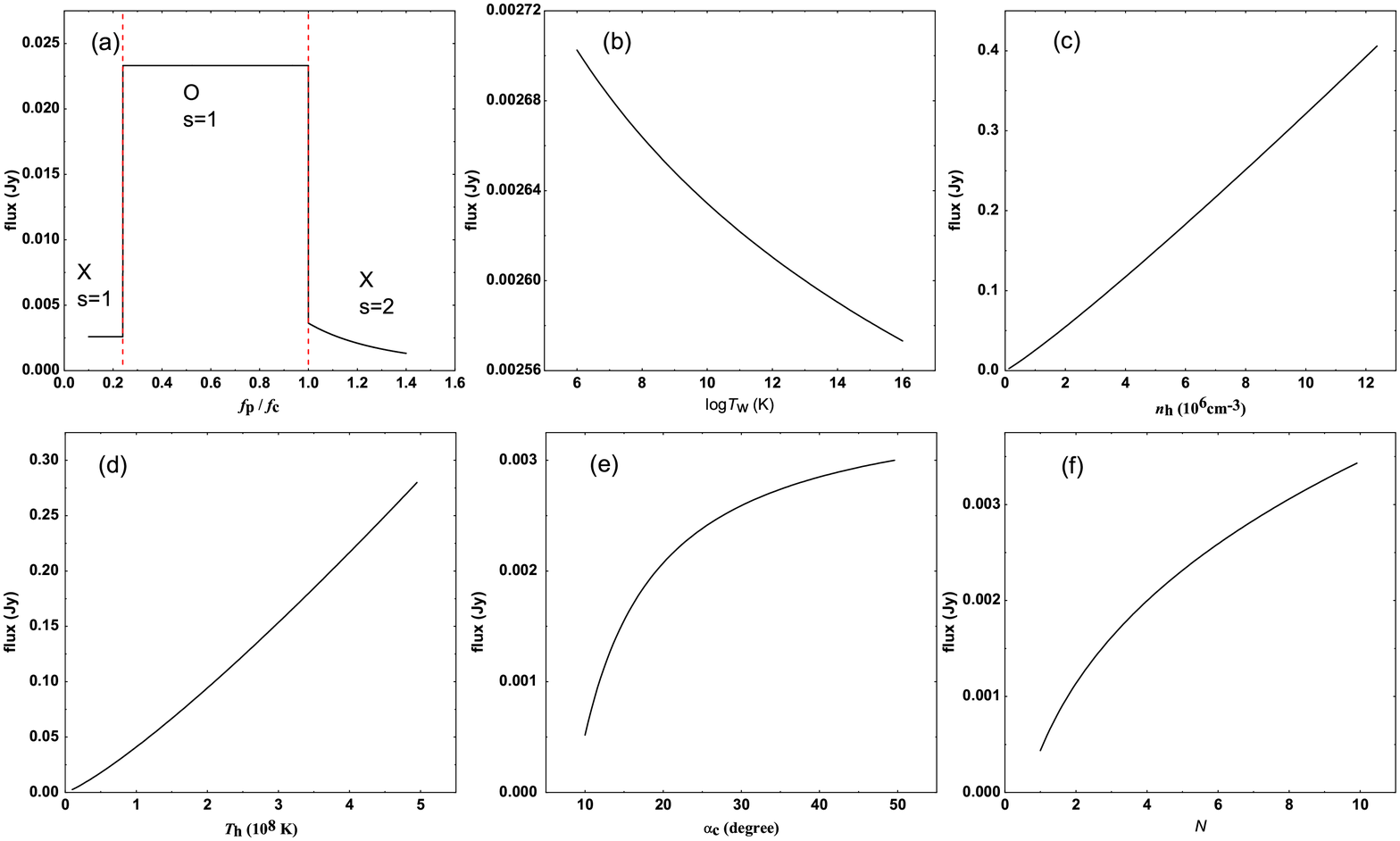}
\caption{Simulated maximum flux density of a radio pulse as a function of the ratio of cyclotron frequency and 
plasma frequency ($f_{\rm p}/f_{\rm c}$), background wave energy ($T_{\rm w}$), density of hot plasma ($n_{\rm h}$), 
temperature of hot plasma ($T_{\rm h}$), loss cone angle ($\alpha_{\rm c}$), loss cone pitch angle distribution 
slope ($N$) are shown in panel (a) to (f) respectively. The parameters for the calculation in this plot are the same as in Fig. \ref{fig_grbt}.}
\label{fig_flux}
\end{figure*}

We can see in panels (c) and (d) in Fig.~\ref{fig_flux}  the variation of the flux density 
($S$) with the number density ($n_{\rm h}$) and temperature ($T_{\rm h}$) of the
hot plasma. Simulations in this section assume standard values for all parameters except $n_{\rm h}$ and $T_{\rm h}$.
Increasing $T_{\rm h}$ from 
$10^{7}$ to $10^{8}$ K, $S$ will increase by $\sim$16 times, while $S$ 
rises by a factor of 12 when $n_{\rm h}$ is one order of magnitude higher. This is 
consistent with the change in the brightness temperature (see Fig. \ref{fig_grbt}).

It is easy to understand that the population of hot electrons is shifted to 
higher velocities where the number of undamped resonance ellipses increases 
and thus the growth rate increases with increase in temperature. On the other hand the 
cold background plasma will negate the loss cone and the growth rate 
will decrease sharply due to the lack of undamped resonance ellipses, if the 
hot electron temperature approaches that of the cold plasma. \citet{Melrose84} 
gives a criterion for effective cyclotron damping by background cold electrons, which is 
$T_{\rm h}/T_{\rm c}<10\sim20$. This value was confirmed in the work of 
\citet{Aschwanden90a}.

\subsection{The effect of cold plasma and the initial wave energy}

The number density $n_{\rm c}$ of the cold plasma can be obtained from the ratio of the plasma 
frequency to the cyclotron frequency ($f_{\rm p}/f_{\rm c}$). Given a magnetic field strength 
($B$=1750 G), $f_{\rm c}$=4.9 GHz. Since the standard value of $f_{\rm p}/f_{\rm c}$=0.1, 
$f_{\rm p}$=490 MHz. This implies $n_{\rm c}=3\times10^{9}\rm cm^{-3}$, the standard 
value adopted in our simulations. Note from Fig. \ref{fig_grbt}, that changing 
$f_{\rm p}/f_{\rm c}$ from say 0.1 to 0.2 makes no difference to the brightness temperature. 

In panel (a) of Fig. \ref{fig_flux} we show the effect of changing $f_{\rm p}/f_{\rm c}$ for a constant 
$B$. We see that the flux density goes up 
a little from $f_{\rm p}/f_{\rm c}$=0.23 to 0.5 ($n_{\rm c}$=1.6$\times10^{10}$ to 
7.4$\times10^{10}$ cm$^{-3}$), and then drops when $f_{\rm p}/f_{\rm c}$ is increased further to 
1.2 ($n_{\rm c}$=4.3$\times10^{11}$ cm$^{-3}$). This result is consistent with the variation 
of the brightness temperature in Fig. \ref{fig_grbt} because the flux density in our simulation 
depends strongly on the brightness temperature.

We note, however, that, although the growth rate declines persistently in Fig. \ref{fig_grbt}, 
the brightness temperature increases steeply by almost one order of magnitude when the 
magneto-ionic mode changes from $X$-mode to $O$-mode at harmonic number $s$=1. This is because 
of the Doppler resonance condition as discussed by \citet{Aschwanden90a}. The resonance ellipses 
covering the unstable portion of the loss-cone distribution form a smaller region of positive 
growth for the $O$-mode, and a  higher wave level results from the same amount of energy 
conversion.

We need to mention that only the dominant magneto-ionic modes, i.e. fundamental ($s$=1) $X$-mode 
when $f_{\rm p}/f_{\rm c}<0.24$, fundamental ($s$=1) $O$-mode when $0.24<f_{\rm p}/f_{\rm c}<1.0$, 
and second harmonic ($s$=2) $X$-mode when $1.0<f_{\rm p}/f_{\rm c}<1.4$, have been taken into account 
in our simulations. When $f_{\rm p}/f_{\rm c}>1.4$, some electrostatic instabilities become 
important rather than the ECM instability, and a condition for ECM emission escaping from plasma is
$f_{\rm p}/f_{\rm c}\ll1$. 
The fundamental ($s$=1) $Z$-mode may be dominant when $f_{\rm p}/f_{\rm c}\approx0.3$ and 
$f_{\rm p}/f_{\rm c}\approx1.1$ \citep{Melrose84,Aschwanden90a}. However, we omit this mode because 
its appearance depends strongly on the ratio $f_{\rm p}/f_{\rm c}$ and another mechanism would be 
needed to convert it to electromagnetic radiation.

When $T_{\rm w}$ (defined as the initial wave energy) results mainly from thermal bremsstrahlung 
it takes a low value of $10^{6}\sim10^{8}$ K. $T_{\rm w}$ may become larger,  in 
the range of $10^{10}\sim10^{16}$ K, due to incoherent gyroresonance or gyrosynchrotron radiation. 
Lower $T_{\rm w}$ can lead to a higher energy conversion efficiency because of the effect on the 
resonance ellipses. The flux density drops slightly with increasing $T_{\rm w}$ as seen 
in panel (b) of Fig. \ref{fig_flux}.

\subsection{The effect of loss-cone parameters}

Panel (e) in Fig. \ref{fig_flux} shows the effect of the loss-cone angle on flux density. 
The rapid increase of flux density when the loss-cone angle ($\alpha_{\rm c}$) changes 
from $10^{\circ}$ to $20^{\circ}$ is the consequence of the significant increase
in the size of the region of velocity space involved. Quasi-linear diffusion influences a large 
number of particles and results in growth and amplification of some wave modes growing. 
On the other hand, when $\alpha_{\rm c}>20^{\circ}$, the positive gradient in the perpendicular direction 
decreases so that the electromagnetic wave energy density can not be amplified effectively. 

The effect of pitch angle distribution slope is similar to that of the loss-cone angle, see 
panel (f) in Fig. \ref{fig_flux}.

\subsection{Determining the environment of the radio-emitting region}
\label{sec_environment}

In this section, we determine the possible range of the different 
parameters. From Table \ref{tab_parameters}, there are at 
least 13 parameters that can influence the radio light curve of TVLM 513;  
we are able to reduce this number, however.
 
Considering the contribution of the degenerate electron gas and the ionic 
Coulomb pressure, the radius is almost a constant around 0.1$R_{\rm \odot}$, 
in the range of 0.08$R_{\rm \odot}$ to 0.11$R_{\rm \odot}$ \citep{Chabrier00}.  

The temperature of cold plasma is more difficult to determine. In this work we assume it is 
around $10^{6}$ K, similar to the typical temperature of the solar corona. On the other hand, 
if the coronal temperature was lower, this would favor wave propagation rather than 
damping. We set the initial wave energy $T_{\rm w}$ at a high level of $10^{14}$ K. On the other hand, 
from Fig. \ref{fig_flux}, the flux density varies only slightly (a factor of 2) when $T_{\rm w}$ 
changes from $10^{10}$ to $10^{15}$ K. We should however note that \citet{Hallinan08}
suggested that the ECM instability may be a viable source of quiescent unpolarized radio 
emission, indistinguishable in temporal and polarization characteristics from
gyrosynchrotron radiation.

For convenience we adopt a 
plausible value of 0.1 for the ratio $u$ of plasma frequency to gyrofrequency ($f_{\rm p}
/f_{\rm c}$), a value that permits unstable growth of the ECMI. If we focus on the observation frequency of 4.9 GHz, the 
magnetic field strength ($B$) has to be around 1750 G and the density of the cold background 
plasma is $\sim$3$\times10^{9}$ cm$^{-3}$. Any change 
of $u$ alone will not affect the maximum flux density for the fundamental $X$-mode and fundamental 
$O$-mode respectively. However, we stress that if we change the magnetic field strength ($B$), 
even keeping $u$ at the same value, the maximum flux density will change because it depends on 
the cyclotron frequency. 

The size of the radio-emitting region $A$ is associated with the time duration of the radio 
pulses. Thus in the case of TVLM 513, we have $A\approx(54.75\Delta t)^{\gamma}$ when the latitude 
of the active region is 30$^{\circ}$. Changing the latitude to 70$^{\circ}$ gives $A\approx(21.62\Delta t)^{\gamma}$.
$\Delta t$ is easily obtained from observations and is a more convenient parameter than $A$. 
$r_{\rm tube}$ is a variable to describe the radius of a flux tube. We expect there are distributions for plasma 
parameters ($T_{\rm h}$, $n_{\rm h}$) and loss-cone parameters ($\alpha_{\rm c}$, $N$) in different flux tubes. 
The variation of different parameters can result in the oscillation of the observed flux density. Smaller values  
for $r_{\rm tube}$ lead to smoother simulated radio light curves (see Fig. \ref{fig_rtubev}). 

So now the free parameters are reduced to 6, i.e. the density $n_{\rm h}$ 
and temperature $T_{\rm h}$ of hot plasma, the angle $\alpha_{\rm c}$ and pitch angle distribution slope $N$ 
of the loss-cone, the radius $r_{\rm tube}$ of one flux tube and the size of the 
emitting region $A$. The functions of the parameters are: $n_{\rm h}$ and 
$T_{\rm h}$ can control the change of flux density dramatically, $\alpha_{\rm c}$ and $N$ 
can control the change of flux density gently, $A$ can control both flux density and time 
duration of the radio pulse which can be constrained by observations, and $r_{\rm tube}$ 
describes the radius of the flux tube in the active region. 

\subsection{Comparison with observations}

\begin{figure*}
\centering
\includegraphics[width=19cm,clip]{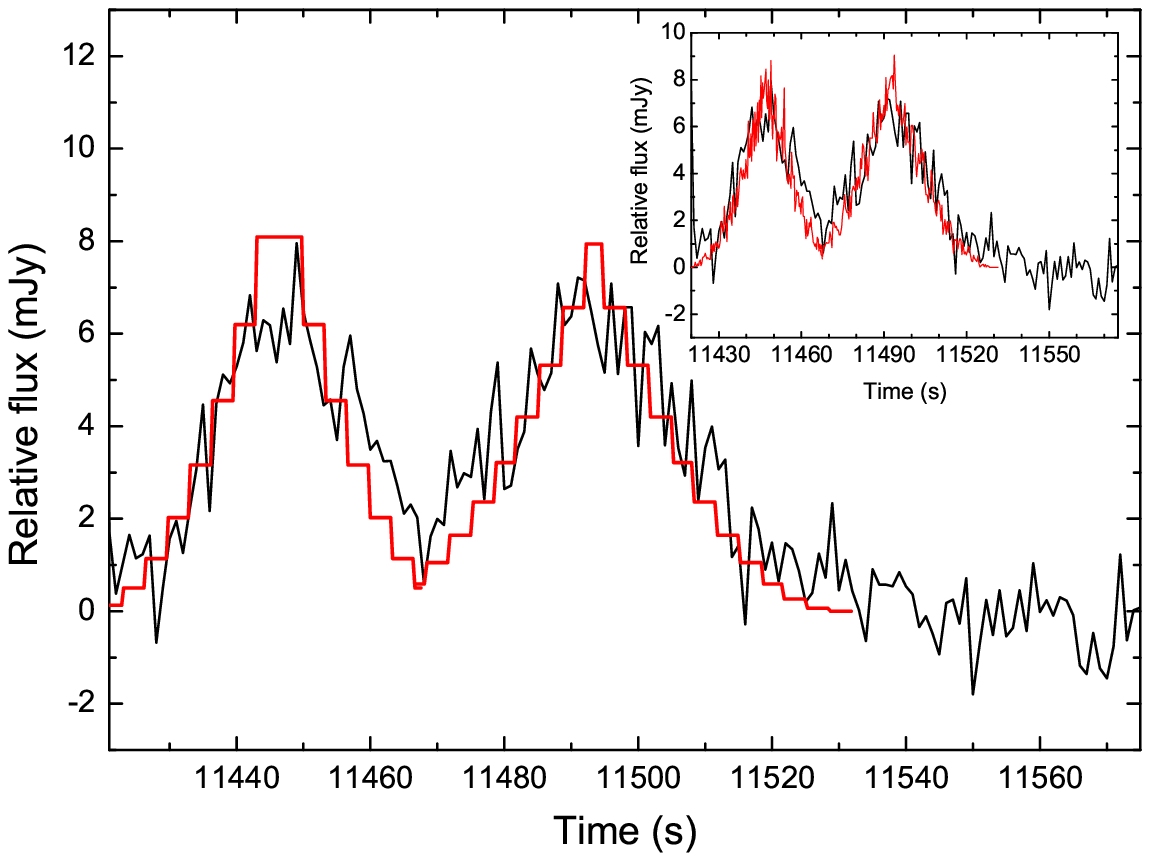}
\caption{Comparison of our simulation with observation taken with the Arecibo telescope by Hallinan et al. (2010).  
Black solid line shows the observation of 2008 May 19 (UT) at 4775 MHz. Red line shows the simulated curve. 
The inset panel shows the simulated light curve using a uniform distribution of the free parameters.}
\label{fig_com}
\end{figure*}

The free parameters can be constrained by making our simulations match up to observations. 
Fig. \ref{fig_com} shows the comparison of our simulation with observations (see Hallinan 
et al. 2010 for further details). The observation is taken from 2008 May 19 (UT) 
at 4725 MHz for the M8.5V dwarf TVLM 513. This means that the magnetic field strength is 
1652 G because of the cyclotron frequency. We adopt $f_{\rm p}/f_{\rm c}=0.1$, implying 
the density of cold plasma in the emitting region is $\sim$2.64$\times10^{9}$ cm$^{-3}$. 
The size of the radio-emitting region is constrained by 
the time duration of the two pulses ($A\approx(54.75\Delta t)^{2}$), implying 1620$\times$1620 km$^{2}$ for the first 
pulse and 2060$\times$2060 km$^{2}$ for the second pulse. The radius of each 
flux tube is taken as 180 km arbitrarily because we are not able to get any information 
on its value from the observed radio light curve. We take $\alpha_{\rm c}=30^{\circ}$ and 
$N=5$ for the first pulse, $\alpha_{\rm c}=35^{\circ}$ and $N=2$ for the second pulse. 
$T_{\rm h}$ and $n_{\rm h}$ are same for the both pulses, equal to $10^{7}$ K and 
$1.25\times10^{5}$ cm$^{-3}$ respectively.

The simulated light curve reproduces the observation. However, this does not mean that 
the above values for the different parameters are unique. In order to compare our simulation 
with the observed two pulses, we take a flat distribution for 
$\alpha_{\rm c}$ and $N$ in the range of $10^{\circ}-50^{\circ}$ and $1-6$ for the both pulses, 
$n_{\rm h}:$ $(1.25-5)\times10^{5}$ cm$^{-3}$, $T_{\rm h}:$ $(1-4.5)\times10^{7}$ K for the first 
pulse and $n_{\rm h}:$ $(1.25-4.5)\times10^{5}$ cm$^{-3}$, $T_{\rm h}:$ $(1-3)\times10^{7}$ K for 
the second pulse (see inset panel in Fig. \ref{fig_com}). We note that the 
size of the emitting region depends on the latitude $\theta$ in our simulation (see \S 
\ref{sec_environment}). Changing the latitude from $30^{\circ}$ (the value for the simulation) 
to $70^{\circ}$ gives a value for the size of 640$\times$640 km$^{2}$ for the first pulse and 
814$\times$814 km$^{2}$ for the second pulse. Other parameters also need to change in order to 
fit the observed light curve. 

The simulation in Fig. \ref{fig_com} is remarkably similar to the observations. Our results 
indicate that the two radio emission regions producing the two pulses are very close, 
which would be consistent with the nature of the cone radiation of ECMI. An important feature 
is that the radio pulses would repeat with the rotation period 
of the UCD. In the case of TVLM 513, the period is $\sim$1.96 hrs. Moreover, we note 
that the decay time of the pulses in some observations is longer than our simulation. 
This is possibly due to the deformation of the radio-emitting region as fast rotation of 
the dwarf can cause the shape of the emitting region to vary from an almost symmetric circle 
to an asymmetric ellipse with a tail. 

\section{Discussion}
\label{sec_discussion}

Our results show that rotation coupled with the ECMI mechanism can account for the flux density 
and polarization of the radio pulses from TVLM 513 successfully. 
We can not exclude the possibility that the depolarization of ECMI could be due to radiation 
transfer of the emission in a neutral atmosphere with lower fractional ionization or 
that inhomogeneous dust clouds \citep{Littlefair08} could 
have an effect on the quiescent emission and the unpolarized components. 
\citet{Hallinan06} suggested that the depolarization 
or mode conversion of the $X$-mode emission occurs in a density cavity, 
as mode conversion of terrestrial kilometric radiation (TKR) from $X$-mode
to $R$-mode in the emitting density cavity (Ergun et al. 2000) may
account for escape of maser emission 
without re-absorption at higher harmonics of the emission frequency.

For the optically thick source, i.e. optical depths $\tau\gg1$, the brightness 
temperature $T_{\rm b}$ can be constrained by Eq. \ref{eq_fluxtube} 
by using the observed flux density. 
In order to see the relation between $T_{\rm b}$ and flux density, we rewrite 
the equation in the form of 
\begin{equation}
T_{\rm b}=5.32\times10^{10}~\rm K\left(\frac{\it{S}_{f}}{4~ \rm mJy}\right)\left(\frac{\it{f}}{4.9 ~\rm GHz}\right)^{-2}
\left(\frac{\it{d}}{10 ~pc}\right)^{2}\left(\frac{\it{R}_{\rm s}}{R_{\rm Jup}}\right)^{-2}
\end{equation}
where $S_{f}$ is the flux density in mJy at the frequency $f$ (GHz), $d$ the 
distance of the radio source from us in pc, $R_{\rm s}$ the size of the emission 
region in Jupiter radii (1 $R_{\rm Jup}$ $\sim$ 0.1 $R_{\odot}$ $\approx$ 7 
$\times$ 10$^{9}$ cm) \citep{Linsky83,Doyle88,Dorman89,Burrows89,Leto00}.
Unfortunately, when we calculate $T_{\rm b}$, we have to assume the size of the 
radio-emitting region. For example, observations show that the flux density of the 
pulses of TVLM 513 is about 4 mJy at $\sim$4.9 GHz. 
\citet{Berger02} obtained the brightness temperature in the range of 
$10^{8}-10^{9}$ K by assuming the size of a corona to be $2-4$ $R_{\rm Jup}$, while 
\citet{Hallinan06} deduced a value of 2.9$\times10^{10}$ K if the size of the region 
is 1 $R_{\rm Jup}$. In our model, $T_{\rm b}\gtrsim5\times10^{10}$ K is about the temperature of 
the quiescent radio emission, assumed to come from a large emission region 
($\lesssim1~R_{\rm Jup}$). For the radio pulses from magnetic loops, the theoretical temperature 
of the coherent ECMI emission can be up to $10^{15}$ K, implying the emission region for the pulses 
is much more compact, e.g. 0.007 $R_{\rm Jup}$. 

The configuration and topology of magnetic field on UCDs remains unclear. In our model, 
we only need a simple dipole poloidal-like field to calculate the flux density and explain 
the high polarisation of the radio pulses. In the case of TVLM 513, multiple bursts of both 
left and right 100\% circularly 
polarized emission in regions of opposite magnetic polarity indicate 
the existence of a dipolar large-scale magnetic field \citep{Hallinan07} or a few 
small active regions with scale $ << 1~R_{\rm Jup}$.

Another important parameter, when combined with the magnetic field strength, is the pitch angle 
$\beta$. Electrons with small $\beta$ less than a critical value $\beta_{\rm c}$, 
i.e. $\beta<\beta_{\rm c}$, precipitating into the dense atmosphere are lost, while the 
electrons with $\beta>\beta_{\rm c}$ will be reflected back to the flux tube to form an anisotropic 
velocity distribution. The value of $\beta_{\rm c}$ depends on the convergence factor which is 
determined by the ratio of magnetic field strengths at the top and the foot-points of the 
flux tube (also called magnetic mirror ratio). For a symmetric flux tube, we have \citep{Dulk85}
\begin{equation}
\label{pitchangle}
\beta_{\rm c}=\rm arcsin(\it B_{\rm top}/B_{\rm foot})^{\rm 1/2}
\end{equation}
where $B_{\rm top}$ and $B_{\rm foot}$ are the magnetic field strengths at the top and 
in the foot-point of the magnetic flux tube, respectively. 
Typical values of $B_{\rm top}/B_{\rm foot}$ in the Sun are in the range of 0.1 to 
0.5. In the simulation, the pitch angle is approximately equal to $30^{\circ}$, 
which means we should have $0.5>\sin \beta_{\rm c}=(B_{\rm top}/B_{\rm foot})^{1/2}$. 
This could give a lower limit on $B_{\rm foot}$.

\begin{figure}
\centering
\includegraphics[width=9cm,clip]{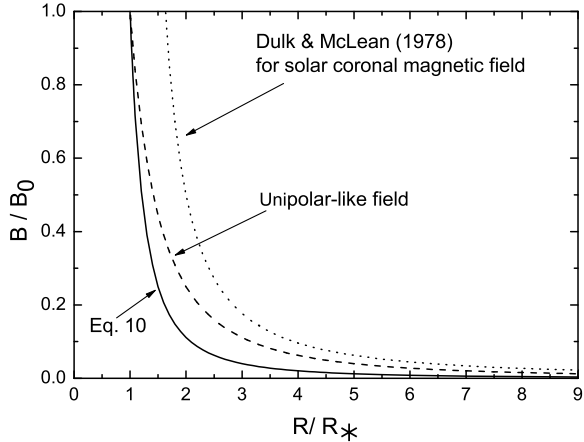}
\caption{Possible magnetic field structures for UCDs. The solid line is for 
Eq. \ref{geometry} which was suggested for a dM star, the dashed line for a unipolar-like field, the dotted 
line for the possible field in the solar coronal region. }
\label{fig_b}
\end{figure}

Assuming the magnetic field is radial, the field strength of a region can be described 
by 
\begin{equation}
\label{geometry}
B=\frac{B_{0}}{4}\left(\frac{R}{R_{\ast}}-\frac{1}{2}\right)^{-2}, ~~~R>R_{\ast}
\end{equation}
where $B_{0}$ is the magnetic field at the photosphere, $R$ the height of the radio emitting 
region measured from the centre of the star and $R_{\ast}$ the photospheric radius. This was chosen as a 
compromise between a unipolar-like field $B=B_{0}(R/R_{\ast})^{-2}$ and the best-fit form 
for solar coronal magnetic fields above active regions $B=0.5[(R/R_{\ast})-1]^{-1.5}$ 
given by \citet{Dulk78} for the range $1.01<R/R_{\ast}<10$. This choice makes the field 
fall off more quickly than either a unipolar magnetic field configuration or that of the solar active region 
magnetic field \citep{Gary81}. Figure \ref{fig_b} shows the dependence of magnetic field strength on height 
of the radio emission region above the photosphere. 

Combining Eqs. \ref{pitchangle} and \ref{geometry}, the radio emission region and therefore 
the magnetosphere can be constrained in the range of 0.56 $R_{\ast}$ to 6.75 $R_{\ast}$. A  
possible position for the radio-emitting region in the radio active UCDs at $\sim$4.9 GHz 
is 1.5 $R_{\ast}$ from their center, 1.06$\times10^{5}$ km in the case of TVLM 513. This implies that the magnetic field 
strength at the photosphere or chromosphere would be as large as 7,000 G. More observations in the optical 
(Zeeman Doppler effect) and infrared bands are needed to constrain 
the magnetic topology and the behavior of the plasma in the magnetic flux tube and to 
determine whether such large field strengths exist on these objects.

In addition, we expect the existence of an enhanced ambient wave energy background by 
gyroresonance turbulence or gyrosynchoroton radiation and some intense events at other wavelengths, 
e.g. optical or X-ray emission, which would occur from the process where hot plasma 
starting from collision-less region collides with the collisional chromosphere or photosphere. 
In the case of X-ray emission, thermal bremsstrahlung emission and inverse Compton scattering could be 
the responsible mechanism since there are hot plasmas with $T_{\rm h}$ up to $10^8$ K and possible low 
energy photons. However, since the Thomson scattering optical depth of the corona of the brown dwarf is $\ll$ 1, 
the contribution of inverse Compton scattering is unlikely to be important.

For the thermal bremsstrahlung emission of ionized hydrogen and helium dominated source, the detected flux density integrated 
over frequency is $S_{\rm X-B}=4.67\times10^{-28}T_{\rm h}^{1/2}n_{\rm h}n_{\rm c}Z^{2}\overline{g}_{\rm B}R_{\rm s}^{3}d^{-2}$, 
where $Z$ is an ion of charge in units $e$ (here we take $Z=1$), $\overline{g}_{\rm B}$ the Gaunt factor (we take 
$\overline{g}_{\rm B}=1.2$, which gives an accuracy of $\lesssim$20\% since $1.1<\overline{g}_{\rm B}<1.5$), 
$R_{\rm s}$ the radius of the source, $d$ distance of the source from us and all quantities are in $cgs$ units \citep{Rybicki79}. 
In the case of TVLM 513, assuming that the X-ray emission comes from the same region as the radio emission, 
we can take the parameters as $T_{\rm h}=10^{7}$ K, $n_{\rm h}=10^{8}$ cm$^{-3}$, $n_{\rm c}=3\times10^{9}$ cm$^{-3}$, 
$R_{\rm s}=2\times10^{8}$ cm, $d=10$ pc$\approx3.1\times10^{19}$ cm, hence we get the $S_{\rm X-B}=4.42\times10^{-21}$ 
erg$\cdot$cm$^{-2}\cdot$s$^{-1}$, giving an X-ray luminosity of $L_{\rm X-B}\approx5.34\times10^{19}$ 
erg$\cdot$s$^{-1}$. The X-ray flux density/luminosity could be underestimated significantly as the X-ray emission 
might be diffuse and from a larger region (perhaps $\sim$10 times) than that of the radio emission,
as is the case for the X-ray emission observed from Jupiter by $Suzaku$ \citep{Ezoe10}.

Interestingly, \citet{Berger08} obtained a marginal detection in X-rays suggesting 
a flux density of $6.3\times10^{-16}$ erg$\cdot$cm$^{-2}\cdot$s$^{-1}$ (luminosity 
$L_{\rm X}=8.5\times10^{24}$erg$\cdot$s$^{-1}$) with mean energy at 0.9 keV. This means 
the temperature of the hot plasma could be slightly lower than $10^{7}$ K. Further 
multi-wavelength observations will help to refine our model and its parameters to understand 
the radio and X-ray emission from these kinds of cool objects. 

\section{Conclusions}
\label{sec_conclusions}

An active region model is applied to the radio emission from a 
cool dwarf, in which the ECMI mechanism is responsible for the radio 
bursts from the magnetic tubes, while the rotation of the dwarf 
can modulate the total observed flux with respect to time. The 
time profile of the radio light curve is in the form of power law in our 
model. Using this model, we can determine the nature (e.g. size, 
temperature, density) of the radio-emitting region plus the magnetic 
topology can be constrained as well. 

In the case of TVLM 513, our model shows the loss-cone electrons have a density 
in the range of $1.25\times10^{5}-5\times10^{5}$ cm$^{-3}$ and temperature 
between $10^{7}$ and $5\times10^{7}$ K. The brightness temperature is typically 
$\sim10^{15}$ K for pulses, $\sim5\times10^{10}$ K for the background emission, 
implying the ECMI mechanism operates in compact region of $\sim$0.007
$R_{\rm Jup}$ if the active region is at 30$^{\circ}$. For an active region closer 
to the pole, e.g. 70$^{\circ}$, the size is $\sim$60\% smaller, implying a higher brightness 
temperature.

The model predicts an enhanced ambient wave energy background and a $\approx$7000 G 
surface magnetic field strength. The theoretical X-ray flux 
density in our model is much smaller than a marginal X-ray observation of TVLM 513, 
which implies a more complicated plasma behavior or magnetic structure on the dwarf.  
Additional multi-wavelength observations are needed to constrain the tentative conclusions and help 
us to improve the understanding of the magnetic field on ultracool dwarfs and to test 
the viability of this model in comparison with others, such as the auroral model.

\begin{acknowledgements}
The Armagh Observatory is supported by a grant from the Northern
Ireland Dept. of Culture Arts and Leisure. GH and AG gratefully acknowledge the 
support of Science Foundation
Ireland (grant No. 07/RFP/PHYF553). AK, SYU \& JGD thank the Leverhulme Trust for support.
AA gratefully acknowledges the support of the Scientific Research Fund of
"St. Kl. Ohridski" University of Sofia (grant No. 80/2009 and 138/2010). ALM \& SYU thank M. 
Aschwanden for providing the quasi-linear diffusion code of ECMI. SYU thanks Gavin Ramsay for 
his comments and also thanks Eamon Scullion for discussions. We also thank the UK Science and Technology 
Facilities Council for support via a Visitor grant. We gratefully thank the referee for his/her 
suggestions and comments. 
\end{acknowledgements}

\bibliographystyle{aa}
\bibliography{15580yu}

\end{document}